\documentclass[structabstract]{aa}

\usepackage{amsmath}
\usepackage[varg]{txfonts}
\usepackage{booktabs}
\usepackage{array}
\usepackage{graphicx}
\usepackage{natbib,twoopt}
\usepackage{upgreek}
\usepackage{color}
\usepackage[breaklinks=true]{hyperref} %% to avoid \citeads line fills
\bibpunct{(}{)}{;}{a}{}{,} %% natbib format like A&A and ApJ
\newcommandtwoopt{\citeads}[3][][]{\href{http://adsabs.harvard.edu/abs/#3}%
{\citealp[#1][#2]{#3}}}
\newcommandtwoopt{\citepads}[3][][]{\href{http://adsabs.harvard.edu/abs/#3}%
{\citep[#1][#2]{#3}}}
\newcommandtwoopt{\citetads}[3][][]{\href{http://adsabs.harvard.edu/abs/#3}%
{\citet[#1][#2]{#3}}}
\newcommandtwoopt{\citeyearads}[3][][]%
{\href{http://adsabs.harvard.edu/abs/#3}{\citeyear[#1][#2]{#3}}} 

\begin{document}

\title{Polarised structures in the radio lobes of \object{B2\,0258+35}}
\subtitle{Evidence of magnetic draping?}
\author{B. Adebahr\inst{1,2}\thanks{adebahr@astro.rub.de}\and M. Brienza\inst{1,3,4}\and R. Morganti\inst{1,3}}
\institute{ASTRON, PO Box 2, 7990 AA Dwingeloo, The Netherlands
\and Astronomisches Institut der Ruhr-Universit\"at Bochum (AIRUB), Universit\"atsstrasse 150, 44780 Bochum, Germany
\and Kapteyn Astronomical Institute, University of Groningen, PO Box 800, 9700 AV, Groningen, The Netherlands
\and INAF - Istituto di Radioastronomia, via Gobetti 101, 40129, Bologna, Italy}

\date{-}
\abstract{The contribution of active galactic nuclei to the magnetisation of the Universe can be constrained by knowing their duty cycles, jet and magnetic field morphologies, and the physical processes dominating their interaction with the surrounding environment.}{The magnetic field morphology and strength of radio lobes of AGN has an influence on the mechanisms for the propagation of cosmic rays into intergalactic space. Using the source B2\,0258+35 we want to investigate the interaction of its radio lobes with the surrounding environment and examine the underlying physical effects.}{Published \ion{H}{i} and radio continuum data at $\uplambda$21\,cm were combined with newly reduced archival Westerbork Radio Synthesis Telescope polarisation data at the same wavelength to investigate the polarised emission in the radio lobes of B2\,0258+35. We assumed energy equipartition between the cosmic rays and the magnetic field to calculate their pressure and investigate the physical processes leading to the detected emission.}{We detected a unique S-shaped diffuse polarised structure. The lobes have a pressure of $p=1.95\pm0.4\cdot 10^{-14}$\,dyn\,cm$^{-2}$. The calculated total magnetic field strengths are low ($B_{eq}=1.21\pm 0.12\,\upmu$G). We observe depolarisation in the northern lobe, which might originate from the  \ion{H}{i}-disc in the foreground. In addition we see an anti-correlation between the pressure and the fractional polarisation along the S-shaped structure. Therefore we consider magnetic draping and magnetic field compression as possible effects that might have created the observed S-shape.}{Our results suggest that magnetic draping can be effectively used to explain the observed polarised structures. This is likely due to the combination of a relatively low magnetic field strength, enabling super-Alfv\'{e}nic motion of the rising lobes (with $M_A=2.47-3.50$), and the coherency of the surrounding magnetic field. Moreover, the draped layer tends to suppress any mixing of the material between the radio lobes and the surrounding environment, but can enhance the mixing and re-acceleration efficiencies inside the lobes, providing an explanation for the average flat spectral index observed in the lobes.}

\maketitle
\section{Introduction}

Radio polarisation studies are powerful probes of the physical processes that shape the evolution of radio-loud active galactic nuclei (AGN) and of the magnetic field structure in these sources. Moreover, they provide us with a unique tool to investigate the properties of the magnetoionic medium that both constitutes and surrounds the source, as well as of the medium that lies between the point of emission and the Earth.

Typical fractions of polarisation in radio galaxies are in the range of 0.1-30\,\% with a median value of 6.2\,\% \citepads{2015ApJ...806...83O}. The highest degrees of polarisation are observed in giant radio galaxies ($>$1 Mpc) with fractions higher than 50\,\% \citepads{1992MNRAS.256..186B}, while most of them reside in the 30\% regime \citepads{2008MNRAS.391..521L}. 

Interferometric studies of radio galaxies over the last decades have revealed that the degree of polarisation is not homogeneous across the radio source, but is rather composed of multiple subcomponents \citepads{1988ARA&A..26...93S}. The degree of polarisation usually increases towards the edges of the lobes as a consequence of magnetic field compression \citepads[e.g.]{1995MNRAS.275L..53E,1997MNRAS.288..859H,2014MNRAS.443.1482H}. Furthermore, polarised structures do not always coincide with total power emission components. In general the polarised brightness distribution shows less symmetry and a more complex distribution than the total intensity emission. 

In double lobed sources a strong asymmetry in the polarisation fraction of the two lobes is often observed. In particular, in double lobed radio sources it is generally found that the jet-side is less depolarised than the counter-jet side. The most well accepted explanation for this trend is the presence of beaming effects \citepads[e.g.]{1988Natur.331..149L,1988Natur.331..147G,1991MNRAS.250..171G}. Indeed, the emission coming from the counter jet travels a longer path within the magnetoionic material that surrounds the source, causing a higher level of depolarisation \citepads{1988Natur.331..149L,1991MNRAS.250..171G,1991MNRAS.250..198G}. While most of the polarisation studies were performed on FRII \citepads{1974MNRAS.167P..31F} radio sources, it has been shown that a polarisation asymmetry also exists for low-luminosity radio galaxies \citepads{1993A&AS...99..407C,1993A&A...267...31P,1997A&A...326..919M} in the B2 sample \citepads{1975A&A....38..209C,1975A&AS...20....1C,1978A&AS...34..341F,1987A&AS...69...57F}.

Spatially resolved observations of radio sources of the FRI-type have shown enhancements of polarised emission on kiloparsec scales along the edges of the radio lobes. However, only a small number of such studies is currently available so that the occurrence of these features has still not been assessed. The sources are usually situated in cluster or group environments \citepads[e.g.]{2011MNRAS.417.2789L,2011MNRAS.413.2525G,2012MNRAS.423.1335G}. All of these studies also revealed diffuse polarised emission encompassing the whole lobe. The common explanation for those structures is the interaction between the jets and the surrounding intergalactic medium (IGM). The plasma from the jet impacts onto a denser surrounding medium and causes magnetic field compression leading to an amplification and ordering of the magnetic field.

In this paper we report the intriguing discovery of two extended polarised structures in the outer lobes of the radio galaxy B2\,0258+35. The structures show a one-sided S-shape that extends along the whole edge of each side of the total power emission of the diffuse radio lobe. Such a morphology has never been observed in any other source before. We present a detailed polarisation study of these structures based on archival observations at 1.4 GHz performed with the Westerbork Synthesis Radio Telescope (WSRT) in 2006.

The radio source B2\,0258+35 consists of a compact steep-spectrum (CSS) radio galaxy \citepads{1995A&A...295..629S,2005A&A...441...89G}, of 3\,kpc size, surrounded by two double lobes with very low surface brightness, which make the total source size $\sim250$\,kpc \citepads{2012A&A...545A..91S,2016AN....337...31B}. According to the recent results of \citetads{2018A&A...618A..45B}, the outer lobes are either still fuelled by the nuclear engine or are a remnant of past jet activity, which switched off not more than probably a few tens of Megayears ago. The radio source is hosted by the SA0 type, \ion{H}{i}-rich \citepads{2010A&A...523A..75S} galaxy NGC\,1167 \citepads{1991S&T....82Q.621D}, located at a distance of 70\,Mpc \citepads{1993AJ....105.1251W}, and is optically classified as a Seyfert 2 type \citepads{1997ApJS..112..315H}. The \ion{H}{i} disc of the galaxy extends to 160\,kpc in diameter with a low surface density ($\leq 2 M_\odot$ pc$^{-2}$). The \ion{H}{i} rotation curve is regular up to a radius of 65\,kpc but shows several signs of recent and ongoing interaction in the very outer parts \citepads{2010A&A...523A..75S}. The galaxy seems to be located in a poor environment with only five other members \citepads{1993A&AS..100...47G}. No diffuse X-ray emission has been detected so far \citepads{2009A&A...500..999A}. Therefore an upper limit to the electron density of the surrounding intergalactic medium is $n_e<3\cdot10^{-4}$cm$^{-3}$ \citepads{1983ApJ...272..439K}.

This paper is structured as follows: Sect. \ref{text_observations} gives an overview of the data used and the reduction of the continuum polarisation data, Sect. \ref{text_results} presents our results, and Sect. \ref{text_polem} analyses possible scenarios to explain the observed polarised structures. The results are discussed in Sect. \ref{text_discussion} and summarised in Sect. \ref{text_summary}. Finally Sect. \ref{text_outlook} presents an outlook on the future of this kind of study.

For this work we used a flat cosmology with the following parameters: $H_0=70$\,km\,s$^{-1}$Mpc$^{-1}$, $\Omega_{\Lambda}=0.7$, $\Omega_M=0.3$. The spectral index $\alpha$ is defined using the convention $S\propto\nu^{-\alpha}$. We assume a redshift of the target of $z=0.0165$ \citepads{1993AJ....105.1251W}, where $1\arcsec$ corresponds to 336\,pc.

\section{Observations}
\label{text_observations}

The polarisation study presented in this paper is based on unpublished data of the radio galaxy \object{B2~0258+35} from the WSRT archive. The data consist of two full 12hr tracks pointing at RA: 03\fh02\fm17\fs8, DEC: +35\degr14\arcmin36\arcsec. The observations were undertaken between 4 and 6 November 2006. Each dataset was recorded over the full bandwidth of 160\,MHz consisting of eight 20-MHz sub-bands with 64 channels each and with centre frequencies equal to 1451, 1433, 1411, 1393, 1371, 1351, 1331, and 1311\,MHz. All four linear correlations (XX, XY, YX, YY) were recorded, so that the data contained full polarisation information. Sources \object{3C48} and \object{3C147} were used as calibrators and were observed before and after the target field in each run, respectively.

\subsection{Data reduction}
\label{text_datareduction}

We performed the data reduction using a combination of the Astronomical Image Processing System (AIPS), the Common Astronomy Software Application package (CASA) V4.7.0 \citepads{2007ASPC..376..127M}, and the Multichannel Image Reconstruction, Image Analysis and Display \citepads[MIRIAD;]{1995ASPC...77..433S} software packages. First, we used AIPS to apply the system temperatures. The data were then converted into the CASA MS-format for radio frequency interference (RFI) inspection. RFI GUI, the graphical front end to RFIconsole \citepads{2010MNRAS.405..155O}, was used to identify failing antennae and solar interference and then to perform an automatic flagging strategy tailored to the WSRT setup. This encompassed the application of a preliminary bandpass calibration to mitigate the steep bandpass slope of the WSRT-receivers, allowing for an easier identification of bad data. 

The actual bandpass calibration was then repeated on the flagged data using the source \object{3C147} followed by gain and polarisation leakage calibration. The solutions were then transferred to the polarised calibrator source \object{3C48} for the derivation of the polarisation angle corrections. Source \object{3C48} has varying polarisation characteristics over the available frequency band and over time, so we  lack a good polarisation model of the source. Therefore, we computed the flux densities of the Stokes parameters $I, Q, U, V$ using the following expressions and the parameters by \citetads{2013ApJS..206...16P} for the frequency range 1.6-30\,GHz:
\begin{align}
I & = I_0 \cdot \left(\frac{\nu}{\nu_0}\right)^{-\alpha}\\
Q & = p \cdot I \cdot \cos\left[{2 \cdot \left( PA_0 + RM \cdot \frac{c}{\nu} \right)^2}\right]\\
U & = p \cdot I \cdot \sin\left[{2 \cdot \left( PA_0 + RM \cdot \frac{c}{\nu} \right)^2}\right]\\
V & = 0 ,
\end{align}
where $\nu_0=1.4603$\,GHz is the reference frequency, $I_0=15.999$\,Jy is the flux density at the reference frequency, $\alpha=-0.8254$ is the spectral index,  $p=0.005$ is the degree of polarisation, $PA_0=2.1293$ is the intrinsic polarisation angle in rad, $RM=-68$\,rad\,m$^{-2}$ is the rotation measure (RM), $c$ is the speed of light, and $\nu$ is the frequency of the channel.
        
Using the relations for the linear feeds of the WSRT,
\begin{equation}
XX = I + Q \quad XY = U + V \quad YX = U - V \quad YY = I - Q,
\end{equation}
we then calculated the flux densities of all four cross-correlations over the observed frequency band.

The above values left us with an error of 20-30$^\circ$ on the polarisation angle (R. Perley priv. comm.) for the final polarisation analysis of the electric field angle values. Therefore, we were not able to reconstruct the absolute polarisation angles. We note that this uncertainty represents a systematic offset of the polarisation angle and may partially affect the degree of polarisation and the rotation measure. This calibration error does not cause any direction- or position-dependent errors between the sources in the field. 

All calibration steps (gain, polarisation leakage, polarisation angle) were derived on a per channel basis. A more detailed description can be found in \citetads{2013A&A...555A..23A}. This calibration approach aims at minimising the effect of the 17\,MHz ripple caused by a standing wave in the WSRT dishes that leads to instrumental polarisation fractions of up to 1.5\,\% \citepads{2008A&A...489...69B} in the centre of the beam. 

After having applied the solutions to the target field, the data were imported into MIRIAD for imaging and self-calibration. The whole self-calibration and imaging process was performed on each individual sub-band. To avoid including  spurious sources in the self-calibration process, we used a mask extracted from the total power image published by \citetads{2012A&A...545A..91S}.

The solution interval as well as the minimal (u,v)-range to include for self-calibration were decreased until the integration time of one minute was reached and all baselines were included. While phase-only solutions were derived in the first iterations, the final steps included both normalised amplitude and phase calibration, as well as solutions for a variance of the XY-phase between solution intervals.

\subsection{Polarisation analysis}

For the final polarisation analysis, we used the RM-synthesis technique \citepads{2005A&A...441.1217B} to mitigate the problem of bandwidth depolarisation. We averaged the data to eight 2.5\,MHz-channels per sub-band and produced 64 cleaned and primary beam-corrected images of Stokes Q and U (one for each channel of each sub-band). This averaging speeded up the imaging, as well as the RM-synthesis processing by an order of magnitude, while preserving the ability to detect structures up to $10160$\,rad/m$^2$, as compression in frequency only affects the maximum observable Faraday depth. To exclude any possible emission coming from the polarised Galactic foreground, we only imaged baselines $\geq450\,\lambda$. 

Images at three different angular resolutions were used as an input for the RM-synthesis to investigate the small-scale structure as well as the large-scale coherency of the magnetic field and faint emission in the lobes. The resolutions used were the highest possible resolution of the image at the lowest frequency (20\farcs9$\times$9\farcs7) and two with circular beams of 30\,\arcsec and 45\,\arcsec. The observational setup results in a resolution in Faraday space of 322.45\,rad/m$^2$ and a maximum observable scale of 74.46\,rad/m$^2$. We sampled the final Faraday cubes between -512\,rad/m$^2$ and +512\,rad/m$^2$ with a sampling interval of 4\,rad/m$^2$.

Since the polarised intensity values are the absolute value of complex numbers, a polarisation bias is introduced. This bias follows non-Gaussian statistics, which is in addition dependent on the distance from the pointing centre due to the primary beam correction. We therefore decided to follow an approximated approach to subtract the polarisation bias. A parabola was fitted to the distance of a pixel from the pointing centre versus the average value along the polarised intensity axis in the Faraday polarised intensity cube. Only emission free areas were considered. We extended this parabola along the Faraday axis creating a polarisation bias cube, which was then subtracted from the polarised intensity cube. The noise level $\sigma$ in the Faraday cubes at all three different resolutions was then estimated by measuring their standard deviation.

We derived the polarised emission map (Fig. \ref{image_PI+TP}) and the rotation measure map (Fig. \ref{image_RM}) at all resolutions by fitting a parabola along the Faraday axis to any values exceeding 4.5$\sigma$. The peak value of the parabola then gives the polarised flux density while the position of the peak in the Faraday spectrum gives the RM value. We also computed a map representing the degree of polarisation (Fig. \ref{image_FP+HI}) by dividing on a pixel-by-pixel basis the polarised intensity map at $45\,\arcsec$ resolution with the total intensity map by \citetads{2012A&A...545A..91S} convolved to the same resolution. We derived the electric field vectors using the Stokes Q- and U-flux densities at the position of the peak in the Faraday spectrum. We did not derive any intrinsic polarisation angles due to their large uncertainties. These errors are mostly due to our broad rotation measure spread function (322.45\,rad/m$^2$) and the uncertainty of the absolute polarisation angle. Nevertheless, the electric field vectors (see Fig. \ref{image_FP+HI}) can be used to illustrate the coherency of the magnetic field.

\section{Results}
\label{text_results}

\begin{figure}
        \resizebox{\hsize}{!}{\includegraphics{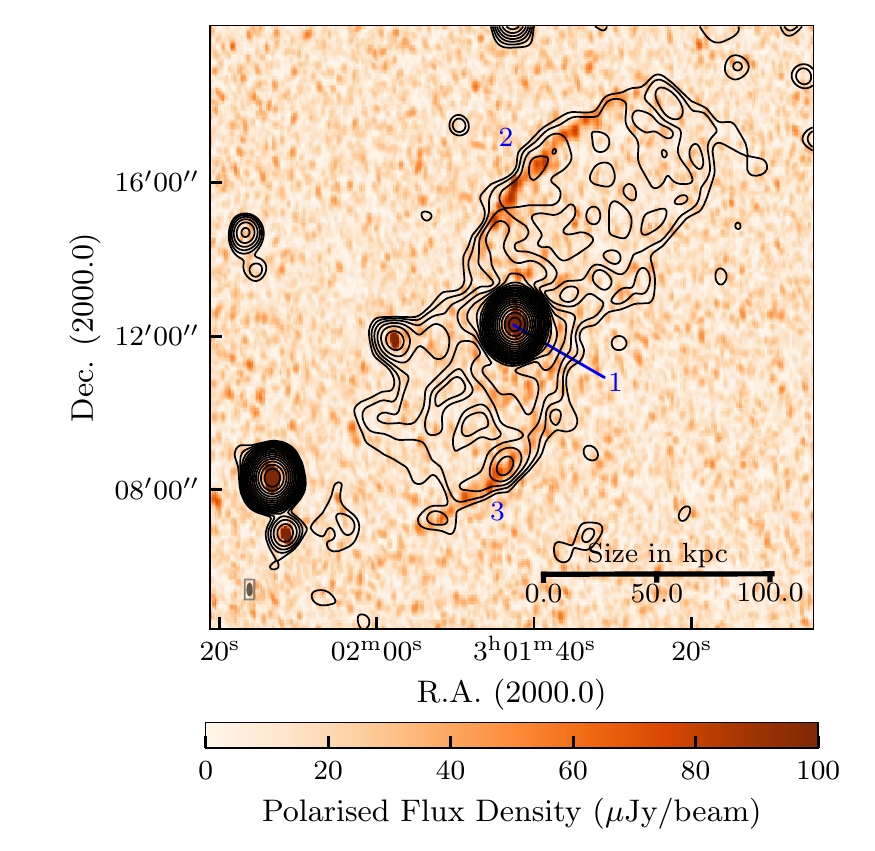}}
        \caption{Total power radio continuum contours at 1.4\,GHz from the WSRT image published in \citetads{2012A&A...545A..91S} overlaid on the polarised intensity image produced at the same wavelength. Contours start at a 3$\sigma$ level of 300\,$\upmu$Jy\,beam$^{-1}$ and increase in powers of 1.5. The resolution of the total power image is 38\farcs8$\times$33\farcs1 and the one from the polarised intensity image is 20\farcs9$\times$9\farcs7, which is shown in the bottom left corner of the figure.}
        \label{image_PI+TP}
\end{figure}

\begin{figure}
        \resizebox{\hsize}{!}{\includegraphics{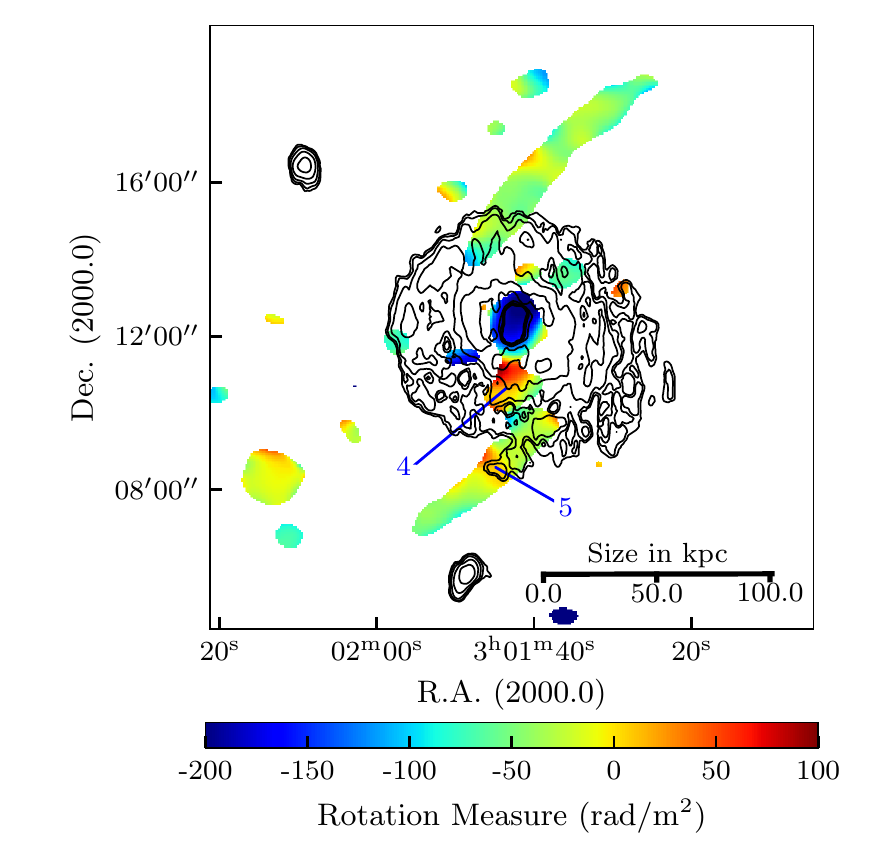}}
        \caption{\ion{H}{i} contours from \citetads{2010A&A...523A..75S} overlaid on the rotation measure map derived from the RM-cube with 45\arcsec\ resolution. The contour levels start at 0.008\,Jy\,beam$^{-1}$km\,s$^{-1}$ and increase in powers of two.}
        \label{image_RM}
\end{figure}

\begin{figure}
        \resizebox{\hsize}{!}{\includegraphics{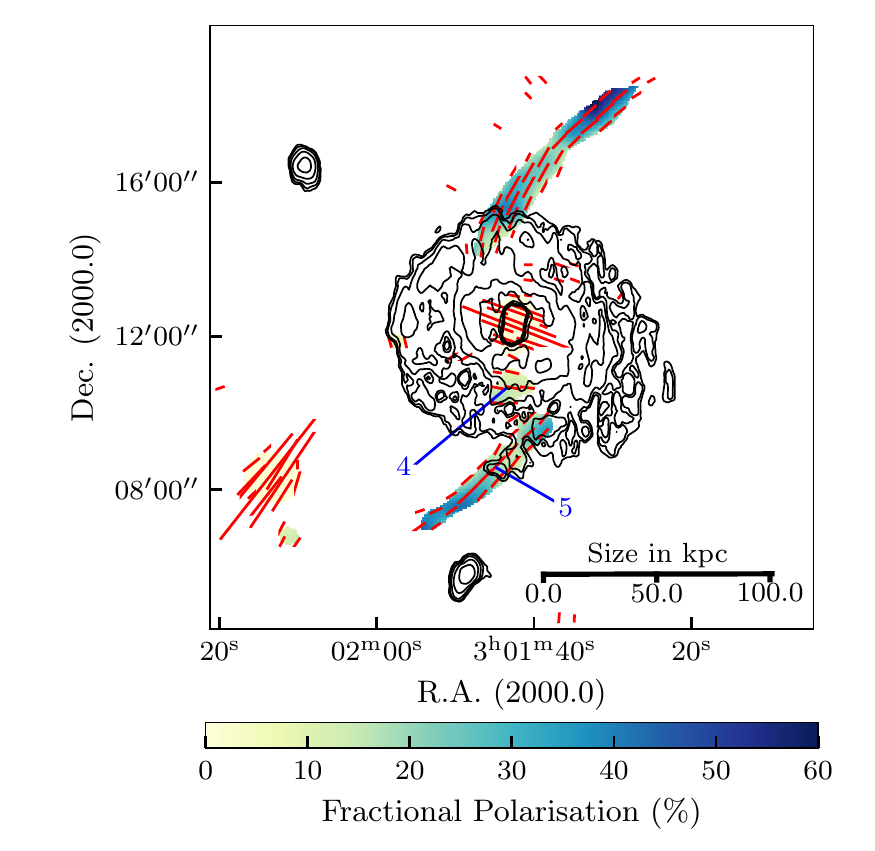}}
        \caption{\ion{H}{i} contours from \citetads{2010A&A...523A..75S} overlaid on the image representing the degree of polarisation at 45\arcsec\ resolution. The contour levels start at  0.008\,Jy\,beam$^{-1}$km\,s$^{-1}$ and increase in powers of two. The red lines represent the electric field vectors where a vector length of 3\arcsec corresponds to 10\,$\upmu$Jy\,beam$^{-1}$ of polarised intensity. These field vectors do not represent the absolute values of electric field vectors (see Sect. \ref{text_datareduction} for further details), but illustrate the coherency of the magnetic field along the S-shaped structures.}
        \label{image_FP+HI}
\end{figure}

To analyse our results we created three different overlays (Fig. 1-3). Figure \ref{image_PI+TP} shows the total intensity contours at 1.4\,GHz from \citetads{2012A&A...545A..91S} with a resolution of 38\farcs8$\times$33\farcs1 on top of our polarised intensity map with a resolution of 20\farcs9$\times$9\farcs7. Figures \ref{image_RM} and \ref{image_FP+HI} show the \ion{H}{i}-emission contours from  \citetads{2010A&A...523A..75S} on top of the derived RM and degree of polarisation maps, respectively. In these figures we marked the most relevant structures (Regions 1-9), which will be described and discussed below.

We have detected polarised emission cospatial with the central compact source (Region 1), as well as along the eastern and western edges of the northern (Region 2) and southern (Region 3) lobes, respectively. The central source (Region 1) shows a polarised flux density of $\rm S_{p,1}=1.74\pm0.11$\,mJy, which corresponds to a degree of polarisation of 0.1\,\%. We note that this polarisation fraction is lower than the polarisation leakage fraction computed for our observation. However, the value of the RM in this region (equal to $\sim-200\,$rad/m$^2$) largely differs from the peak of the instrumental polarisation (equal to $0\,$rad/m$^2$), suggesting that the polarisation signal is real.

The extended polarised structures (Regions 2 and 3) show a unique elongated morphology reminiscent of an S-shape, which has never been observed before. They are co-located with the strongest total power emission and show an integrated polarised flux density equal to $\rm S_{p,2}$=2.44\,mJy and $\rm S_{p,3}$=1.49\,mJy. Both (Regions 2 and 3) extend all along the radio lobes for up to $\sim$100\,kpc but remain spatially unresolved radially, even in the Faraday cubes with the highest angular resolution of $14\arcsec$ (4.75\,kpc).

The degree of polarisation of both structures increases moving from the central source outwards. In Region 2 the degree of polarisation varies from $20\pm2\,\%$ to $45\pm$3\,\% and in Region 3 it varies from $20\pm2\,\%$ to $60\pm$4\,\%. Both regions (2 and 3) show negative RM values (Fig. \ref{image_RM}) with a mean value of $\sim-45$\,rad/m$^2$. Only in Regions 4 and 5 do we observe positive RM values equal to $RM_{4}=\sim50$\,rad/m$^2$ and $RM_{5}=\sim90$\,rad/m$^2$ and much lower values for the degree of polarisation equal to $\sim13\pm$2\,\%. As the source B2~0258+35 is most likely located in a poor environment where we expect electron densities of the order of $n_e\approx3\cdot10^{-4}$cm$^{-3}$ \citepads{1983ApJ...272..439K}, we are led to believe that the rotation measure is produced by a medium co-located with the source itself most likely generated during earlier periods of activity.

It is remarkable how ordered the electric field vectors in Fig. \ref{image_FP+HI} are. The whole diffuse structure shows vectors following its elongation with average values of $PA\approx40\degr$ and a standard deviation of only $\Delta PA\approx10\degr$. Even the slight bending of the structures towards the end of the lobe is noticeable. The coherency of the vectors confirms that we do not encounter any position-dependent effects from the polarisation calibration. Unfortunately, as mentioned above, we cannot derive the de-rotated magnetic field vectors reliably and therefore are limited to an analysis of the relative values and coherency of the emission.

We note that the southern polarised emission region extends all the way from the nucleus until the end of the radio lobe, while the northern one shows a gap between the nucleus and the extended radio lobe. Comparing the polarised morphology with the \ion{H}{i} data from \citetads{2010A&A...523A..75S} (see Figs. \ref{image_RM} and \ref{image_FP+HI}), we might be able to attribute this lag of emission to \ion{H}{i} from the disc of the galaxy in the foreground. We will further discuss this topic in Sect.~\ref{text_discussion}.

\section{Origin of the polarised emission in the outer lobes}
\label{text_polem}

In this section we investigate two different physical mechanisms that may be the origin of the polarisation properties observed in B2~0258+35: magnetic field compression and magnetic draping. Both mechanisms require an interaction between the extended lobes' plasma and a second medium. We mainly consider here the second medium to be the surrounding intergalactic medium (IGM), but we also mention the possibility of an interaction with plasma associated to a restarted jet from previous cycles of activity within the extended lobes in Sect. \ref{text_discussion}.

\subsection{Magnetic field compression}
\label{text_magcompr}

Compression in the shocks of super-sonically expanding lobes into the IGM and a consequent magnetic field compression may occur if the radio lobes propagate into a region of higher density. This has been observed in radio galaxies located in galaxy clusters or groups. These kinds of compression regions are usually visible in the X-ray regime \citepads[e.g.]{2018ApJ...855...71S,2009MNRAS.395.1999C,2007ARA&A..45..117M,2004ApJ...607..800B}. X-ray observations of B2~0258+35 by \citetads{2009A&A...500..999A} only show emission of the central source, but these data are shallow and deeper observations are needed to measure the density of the IGM. In the following we investigate the viability of a compression scenario.

To investigate whether the elongated polarised structures observed in B2~0258+35 (Regions 2 and 3) are the result of lobe plasma compression, we derive spatially separated pressure values over the whole extends of the lobes using the magnetic field pressure as the basis.

\begin{figure}[t!]
        \resizebox{\hsize}{!}{\includegraphics{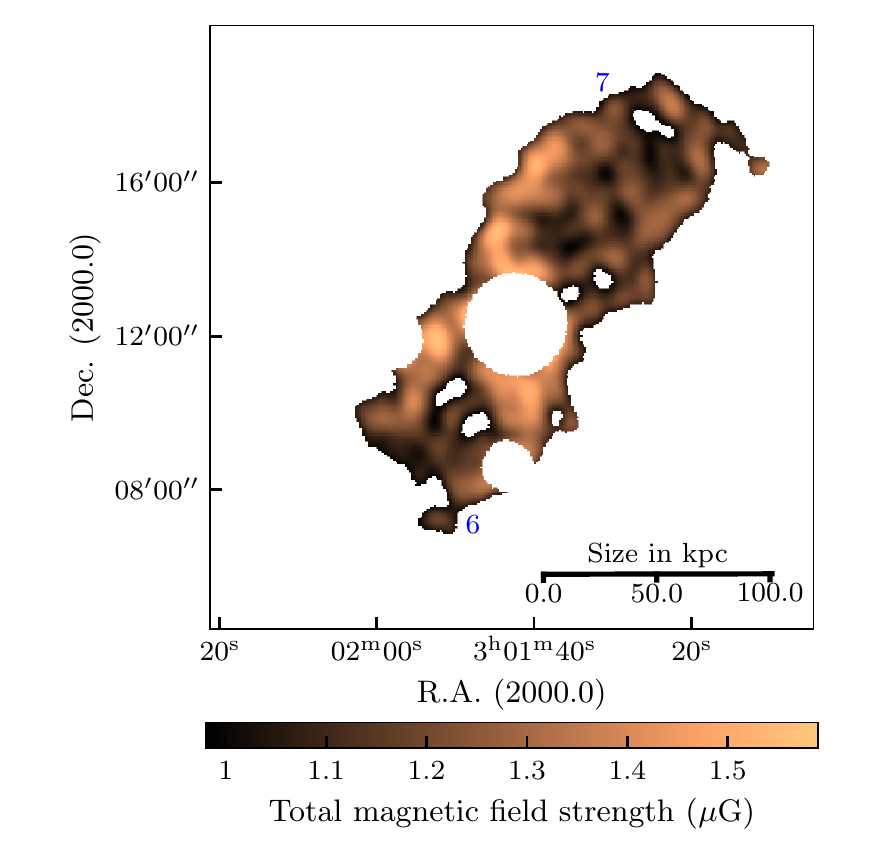}}
        \caption{Total magnetic field strength calculated from energy equipartition \citepads{2006LNP...693...39W} with the assumption of a cylindrical geometry for the radio lobes as explained in Sect. \ref{text_magcompr}.}
        \label{image_Btot}
\end{figure}

We followed the recipe described in \citetads{2000A&AS..146..293S} to compute the pressure inside the lobes. We first calculated the equipartition magnetic field strength $B_{eq}$ using the equations from \citetads{2006LNP...693...39W}. We assumed a power-law particle distribution with Lorentz factors between $\gamma_{min}=10$ and $\gamma_{max}=10^6$. A proton-to-electron ratio of $K_0=1$ and a constant spectral index of $\alpha=0.6$ was used, which corresponds to the average value of the two lobes from Brienza et al. submitted. We assumed a cylindrical geometry with a diameter of 130\,kpc, which is approximately equal to the lobe width computed using the $3\sigma$-contours in the total intensity map. The calculation of the magnetic field strength was executed for each pixel by calculating the encompassed volume of one pixel along the line of sight as an input. We blanked any pixels in the map that we could identify as background point sources, as well as the region corresponding to the emission from the strong central nucleus. The resulting map is shown in Fig. \ref{image_Btot}. Using this approach we found magnetic field values across the lobes in the range $0.98-1.59\,\upmu$G with an average value of $1.21\pm 0.12\,\upmu$G.

This value confirms the value calculated by \citetads{2018A&A...618A..45B} of $1\,\upmu$G, who used the integrated flux density of the entire lobes to calculate the magnetic field strength. We will discuss the consequences for different $K_0$-ratios and their impacts on our calculations in Sect. \ref{text_discussion}. 

From the magnetic field strength we can directly calculate the energy density $u_{eq}$ of the lobes using the standard textbook formula

\begin{equation}
u_{eq} = \frac{B_{eq}^2}{8\pi}
.\end{equation}

In a relativistic plasma, the pressure $p$ is directly related to the energy density as

\begin{equation}
p = \frac{1}{3} u_{eq}.
\end{equation}

\begin{figure}
        \resizebox{\hsize}{!}{\includegraphics{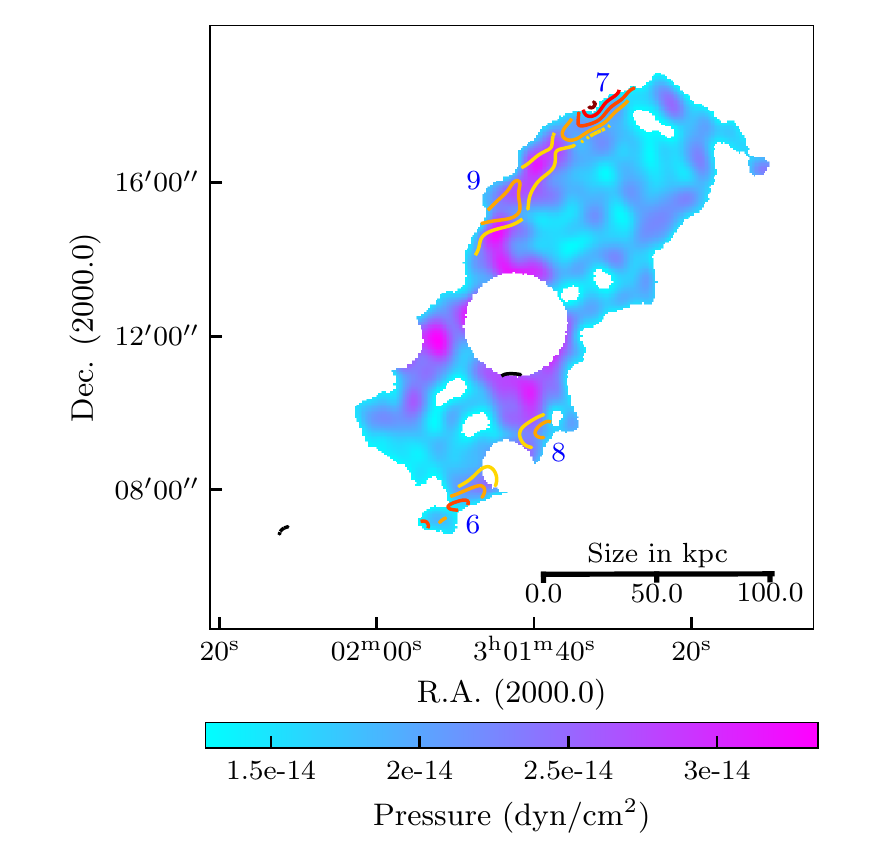}}
        \caption{Pressure within the radio lobes calculated from the energy density of the magnetic field. Contours show the degree of polarisation at 10 (yellow), 20, 30, 40, 50, and 60 (red) percent levels.}
        \label{image_Bpress}
\end{figure}

From the map of the magnetic field strength we can therefore derive the pressure distribution inside the lobes as shown in Fig. \ref{image_Bpress}. The pressure values vary in the range $p=1.28\cdot 10^{-14}-3.34\cdot 10^{-14}$\,dyn\,cm$^{-2}$. These values are similar to those derived by \citetads{2000A&AS..146..293S} for a typical radio galaxy in a group environment.

We note that the magnetic field is strongest along the polarised emission structures (Regions 2 and 3). The highest values are observed in the diffuse emission close to the central compact source with $B_{eq}=1.5\,\upmu$G and $p=3\cdot 10^{-14}$\,dyn\,cm$^{-2}$. In the rest of the lobes the values are as low as $B_{eq}=1.1\,\upmu$G and $p=1.7\cdot 10^{-14}$\,dyn\,cm$^{-2}$. 

Taking a closer look at Regions 2 and 3, where the polarised structures were detected, we see an anti-correlation with the degree of polarisation distribution along these structures. The overall magnetic field strength and pressures are high ($B_{eq}=1.4\,\upmu$G and $p=2.8\cdot 10^{-14}$\,dyn\,cm$^{-2}$), but higher degrees of polarisation go along with lower magnetic field strength and pressures. At the location of the highest degrees of polarisation (Regions 6 and 7) with values of $FP\geq35\%,$ the magnetic field strength and pressure only reach values comparable to those measured in the unpolarised regions of the lobes of $B_{eq}=1.1\,\upmu$G and $p=2\cdot 10^{-14}$\,dyn\,cm$^{-2}$ respectively (Regions 6 and 7).
We use the relation

\begin{equation}
p_{IGM} = 1.0\cdot10^{-14}\,(1+z)^5\,\textnormal{dyn\,cm}^{-2}
\end{equation}

from \citetads{1993MNRAS.260..908S}, who assume an isotropically expanding Universe, to acquire a first order estimate for a lower limit of the pressure in the surrounding medium of the source outside of the radio lobes. This results in a value of $p_{IGM} = 1.08\cdot 10^{-14}$\,dyn\,cm$^{-2}$. Since B2~0258+35 is located in a poor environment and the lobes are probably sitting well outside the galactic coronal halo, this low value in comparison to galaxies in cluster environments, where the pressure reaches a few-tens of $10^{-12}$\,dyn\,cm$^{-2}$ \citepads{1988A&A...189...11M}, seems realistic. However, since we do not have any observational information on the surrounding environment from for example X-ray observations, our calculated value is only a lower limit and should be counted as an order of magnitude estimate.

The lower value for the radio lobes in comparison to the IGM pressure is not surprising since radio galaxies are not expected to be strongly overpressured with respect to the environment and are not expected to drive strong shocks if only the equipartition magnetic pressure is taken into account \citepads{2005ApJ...635..894F,2005ApJ...625L...9N,2005ApJ...628..629N,2006ApJ...644L...9W}. An anti-correlation of the degree of polarisation with pressure, as we see in our data, is not expected from a compression scenario. Higher polarisation degrees need pressure gradients along the compression front to amplify the regular magnetic field. Therefore we investigate magnetic draping as an alternative effect to create the observed structures in the following subsection.

We mention here that these numbers strongly depend on the choice of $K_0$, which has been investigated by different authors. Values of $K_0>0$ were already discussed by \citetads{1978MNRAS.182..443B} and are needed for models of large-scale acceleration processes. Recent publications have also discussed the entrainment of particles from the surrounding medium by radio jets or lobes \citepads{2018MNRAS.476.1614C,2014MNRAS.443.1482H,2008MNRAS.386.1709C}. Such a scenario would raise the $K_0$ value, due to more protons than electrons being entrained, and lead to a higher magnetic field strength. For example, $K_0=100$ would raise the calculated values of the magnetic field strength by a factor of $\sim3.5$ and the pressure by a factor $\sim4$. Other uncertainties in the calculations come from the simple assumption of a cylindrical geometry and a uniform spectral index.

\subsection{Magnetic draping}
\label{text_magdrap}

The physical effect of magnetic draping is significantly different from magnetic field compression. While compression orders and amplifies the magnetic field along the whole path travelled by the shock front, draped field lines are folded around the compression front and move back to their original configuration after the front has passed \citepads{2008ApJ...677..993D}. In a steady state scenario, the rate at which the lines pile up at the front of the draping layer is the same as the rate at which they are advected over the surface and eventually leave it \citepads{2018arXiv180605679E}. The prerequisites for magnetic draping are a sufficiently large magnetic coherence scale $\lambda_B$ and super-Alfv\'{e}nic motion of the propagating medium. This super-Alfv\'{e}nic motion results in a complete isolation of the two interacting magnetic fields from each other and therefore prohibits any magnetic mixing of the two media. This is a significant difference to the magnetic field compression scenario where compressed layers can be mixed and lead to instabilities along the interaction regions.

\citetads{2006MNRAS.373...73L} pointed out that radio lobes moving through the intracluster medium at super-Alfv\'{e}nic speeds can lead magnetic draping to occur and form a thin magnetised boundary layer. Recent simulations by \citetads{2018arXiv180605679E} could even reproduce and resolve magnetic draping layers in simulations of magnetised jets in galaxy clusters.

In the case of B2~0258+35 we first consider this boundary layer to develop between the radio lobe and the IGM and will discuss a possible interaction of a propagating jet in a fossil radio lobe in Sect. \ref{text_discussion}.
A large-scale coherent magnetic field can be assumed for the surrounding medium as well as for the jet. Indeed, \citetads{2008A&A...483..699G,2010A&A...514A..50G} showed that the magnetic field in the IGM and intracluster medium (ICM) is regular on scales of tens of kiloparsecs. This is also observed in radio lobes, where there are high degrees of polarisation and coherent magnetic field morphologies \citepads[e.g.]{1994A&A...283..729K,1996A&A...306..708S,2011MNRAS.413.2525G}.

In order to test whether magnetic field draping is a viable explanation for the polarised structures observed in B2\,0258+35, we evaluate the value of $\lambda_B$ in this source. For this we use two approaches. As a first method, we calculate the Alfv\'{e}n speed $v_A$ using the standard textbook equation

\begin{equation}
v_A = 2.18\,{n_e}^{-\frac{1}{2}}B\,\textnormal{km\,s}^{-1}
\label{Eq_Alfven_speed}
,\end{equation}

with $n_e$ being the electron density in particles per cm$^{-3}$ and $B$ the magnetic field strength in $\upmu$G. Using an average magnetic field strength of $B_{eq}=1.4\,\upmu$G and a typical thermal electron density of $n_e=10^{-4}$\,cm$^{-3}$ \citepads{2013ApJ...764..162O}, we obtain a value of $v_A=305$\,km\,s$^{-1}$. Assuming an approximate age of the bubble of $\sim$110\,Myr \citepads{2018A&A...618A..45B} and a height $H=120$\,kpc, we get an average expansion speed of $v_{exp}\approx$ 1067\,km s$^{-1}$. By combining the two speeds computed above, $v_A$ and $v_{exp}$, we get a super-Alfv\'{e}nic scenario with a Mach number of $M_A\leq3.5$. We note that if we assume a higher electron density closer to the central source, $v_A$ would reduce and therefore $M_A$ would be higher. Using the equation from \citetads{2008ApJ...677..993D},

\begin{equation}
\lambda_B\gtrapprox R/M_A,
\end{equation}

where $R$ is the curvature radius at the stagnation point of the draping layer and $M_A$ the Alfv\'{e}nic Mach number, with $R=50\,$kpc, we obtain a magnetic coherence scale equal to $\lambda_B=14$\,kpc.

As a second method, we calculate a lower limit to $M_A$ using the relation between the lobe radius $R$, the thickness of the boundary layer $\Delta r,$ and $M_A$ as proposed by \citepads{2006MNRAS.373...73L}:

\begin{equation}
\frac{\Delta r}{R} \propto \frac{1}{{M_A}^3}
.\end{equation}

By assuming $R=2.5'$ (50\,kpc) and $\Delta r=10''$ (3.4\,kpc), as measured from the highest resolution polarised intensity map, we can estimate an Alfv\'{e}nic Mach number of $M_A=2.47$ or an expansion speed of $v_{exp}=$ 753\,km s$^{-1}$ (assuming  $v_A=305$\,km\,s$^{-1}$). We derive therefore in this way a magnetic coherence scale of $\lambda_B=20$\,kpc.

Both methods presented above produce results that meet the requirement $\lambda_B\gtrapprox R/M_A$ necessary for magnetic draping to develop. This suggests that draping can develop between the expanding lobes and the surrounding IGM of B2\,0258+35.

\section{Discussion}
\label{text_discussion}

In the previous section we have presented the unique polarisation morphology of the source B2~0258+35 and investigated two possible physical mechanisms to explain its origin: magnetic field compression and magnetic draping. Our results seem to favour the magnetic draping mechanism. In this section we discuss our results on both mechanisms in the context of previous literature studies. 
 
Magnetic draping is mostly known from solar system physics \citepads[e.g.]{2018GeoRL..45.3356F,2017ApJ...839L..12O,2015GeoRL..42.1640L,2015ApJ...805..153I}, but has become relevant over the last couple of years to explain magnetic field morphologies of merging galaxy clusters, such as the famous bullet cluster \citepads{2012MNRAS.423..236R}, and motions of galaxies in cluster environments \citepads{2014ApJ...784...75R}. \citetads{2010NatPh...6..520P} were even able to use the signatures of the draped layers to investigate the orientation and scale length of the magnetic field inside the intracluster medium of the Virgo cluster. \citetads{2012ApJ...761..185Y} successfully reproduced the morphology of the Fermi-bubbles and concluded that the suppression of magnetohydrodynamic (MHD) instabilities can be explained by a draped magnetic field layer. Later, \citetads{2013MNRAS.436.2734Y} found increased degrees of polarisation inside the draping layers. Polarisation analysis of four nearby AGN by \citetads{2011MNRAS.413.2525G} showed RM bands orthogonal to the source axis within their lobes, which could only be explained by magnetic draping. \citetads{2008ApJ...677..993D} simulated magnetic fields over rising radio bubbles and predicted a similar appearance of a draped layer as we observed for B2~0258+35 under certain viewing angles. Recently \citetads{2018arXiv180605679E} were able to recover and resolve draping layers in simulations of magnetic jets interacting with the surrounding environment. Despite the variety of extragalactic sources described above in which magnetic draping has been suggested to play a role, none appears to be similar to the radio galaxy B2\,0258+35 studied in this paper.

It is interesting to note that the non-detection of diffuse polarised emission in the northern radio coincides with the border of the \ion{H}{i}-disc. This could be caused by depolarisation effects. We know that at 1.4\,GHz discs of galaxies are not transparent to polarised emission and therefore the origin of the recovered polarised signals is most often related to the front sides of their discs or halos \citepads{2009A&A...503..409H}. The reason for this is a depolarisation of the polarised emission of the backside of the galaxy due to the small-scale turbulent field in the star-forming regions between the observer and the emitting region. The same effect could depolarise the emission of the northern jet if it is situated behind the disc of NGC\,1167 while the southern one is situated in front of it.

An alternative option would be a scenario where polarised emission in the rest of the radio lobes is too faint to be detectable with our observations. Such a morphology would support a magnetic draping scenario. Polarised emission is only visible where draping compresses the ordered magnetic field lines and the intensity of polarised emission rises.

In addition to the missing polarisation in the northern lobe and the overall disc of the galaxy, we can see two small regions of lower degrees of polarisation in the filamentary structures (Regions 8 and 9). We can identify two optical counterparts in Sloan Digital Sky Survey (SDSS) images \citepads{2017ApJS..233...25A}. These sources could be background radio galaxies. Depolarisation can occur if these sources emit polarised emission where the electric field vector of this emission reaches the polarised structures in the radio lobes of B2\,0258+35 with a different angle. In order to disentangle the two components and recover the full polarised signal, the resolution in Faraday space needs to be higher than the separation of the two components. This is not necessarily given with our resolution of 322.45\,rad/m$^2$.

For Region 5 we can see \ion{H}{i} emission on top of a region with lower degrees of polarisation of $\sim15\%$ in comparison to adjacent regions with $25-30\%$ degrees of polarisation. The structure shows a slightly offset mean velocity of the \ion{H}{i}-gas compared to the rest of this side of the disc \citepads{2010A&A...523A..75S}. This could hint at a tidal dwarf galaxy or an \ion{H}{i}-stream on the line of sight. Under the assumption that these regions usually host a more turbulent magnetic field than the rest of the halo of a galaxy, emission can become partly depolarised.

Unique in B2~0258+35 is the S-shaped structure of the polarised emission in the radio lobes. This S-shape could be the result of jet precession or a permanent change of the axis of the jet. Since the polarised emission is not present along the whole edge of the lobe, the precession period would need to be shorter than the jet's active phase so that the jet plasma was preferably ejected in one direction. This could also mean that the observed magnetic field is intrinsic to the jet itself rather than being the result of an interaction of the jet with the IGM. A longitudinal alignment of the field in the inner part could transform into a radial and toroidal field, which is then perpendicular in projection. Such a transformation has been seen in decelerating jets on scales of tens of kiloparsecs and is not unexpected since in adiabatic magnetic field models, the longitudinal component decreases faster when the jet expands.

Another option would be a change of the direction of the jet due to interaction with gas in the central regions. Evidence that such an interaction could be ongoing comes from the morphology of the kiloparsec-scale jet and as well as from the  \ion{H}{i} (in absorption) and molecular gas in the centre (Murthy et al. in prep).

Although there are no detections in other radio galaxies of polarised structures similar to those observed in B2~0258+35, we discuss here two sources with some comparable characteristics: NGC\,3998 \citepads{2016A&A...592A..94F} and Mrk\,6 \citepads{2006ApJ...652..177K}. Both objects show an edge brightening of the large-scale radio lobes along their edges, with the lobes of NGC\,3998 even being S-shaped. While  \citetads{2016A&A...592A..94F} did not have any full polarisation information, Mrk\,6 shows high degrees of polarisation at the edges of the lobes of up to $\sim50\%$ and less than $1\,\%$ near the centre. High resolution images of Mrk\,6 \citepads{2006ApJ...652..177K} show radio emission associated with the nucleus on three scales, the mentioned large-scale bubbles ($\sim4.5$\,kpc\,$\times\,7.5$\,kpc), a pair of inner bubbles ($\sim1.5$\,kpc\,$\times\,1.5$\,kpc), and a 1\,kpc radio jet extending into a different direction. For both objects the authors claimed that the difference in the alignment of the jet axis was caused by a precession of the central AGN.

Due to these similarities and the fact that high-resolution observations by \citetads{2005A&A...441...89G} show a small-scale 0.9\,kpc sized jet associated with the nucleus pointing in a different direction from the large-scale emission of B2~0258+35, we also discuss the possibility of a restarted jet scenario. Such an episodic event would lead to a new jet propagating into a fossil one. In this scenario the fossil lobes already generated an ordered magnetic field into which the new ejecta propagate. This could be one reason why the outer lobes have not developed ultra-steep spectral indices with $\alpha\gtrapprox1.2$ up to 6.6\,GHz \citepads{2018A&A...618A..45B}, as expected for very old remnants. Indeed the restarting jets may be fuelling the extended lobes with some fresh particles preventing the spectral steepening.

We note, moreover, that a consequence of a draping layer is the confinement of the particles to the lobes. A draping layer inhibits any particle transport through it, out of the bubble via cosmic-ray diffusion \citepads{2008MNRAS.383.1359R}. Therefore particles would only be able to exit the lobes towards the opposite side of the leading edge of the lobes. This would increase the intensity of the mixing of differently aged plasma components inside the lobes and therefore suppress spatial gradients in the spectral index. In addition it would slow down a steepening of the spectral index due to a higher efficiency of diffusive re-acceleration mechanisms inside the lobes \citepads{2018JPlPh..84c7101C}. In addition, a draping layer stabilises a rising bubble against Kelvin-Helmholtz or Rayleigh-Taylor type instabilities \citepads{2007ApJ...670..221D}. This can explain why in our case the polarised structure can still maintain its coherency and why such diffuse lobes have not rapidly expanded into the external environment.

The reason for magnetic draping being an effective physical process to create the observed emission in our target galaxy is most likely the combination of a relatively low magnetic field strength enabling super-Alfv\'{e}nic motion of a rising bubble and the coherency of the surrounding magnetic field being situated in the IGM or generated by an earlier activity cycle of the source. 

\section{Summary}
\label{text_summary}

In order to investigate the polarisation properties of the radio galaxy B2\,0258+35 we reduced and analysed archival WSRT data at 1.4\,GHz. The source B2\,0258+35 is hosted by the SA0 type galaxy NGC\,1167. While earlier observations already showed faint  radio lobe emission with a total projected size of $\sim250$\,kpc with a strong compact source in the centre, we could detect a very thin elongated polarised structure along the eastern side of the northern lobe and on the western side of the southern one. This S-shaped structure shows unusual high polarisation degrees of up to 60\,\% and up to 45\,\% in the northern and southern lobes, respectively.

By computing the magnetic field and pressure distribution within the radio lobes, we have investigated the origin of these structures. In particular, we have considered two physical mechanisms, plasma compression and magnetic draping. We used the assumption of energy equipartition to estimate a magnetic field strength of $0.98-1.59\,\upmu$G and pressures of  $p=1.28\cdot 10^{-14}-3.34\cdot 10^{-14}$\,dyn\,cm$^{-2}$ in the radio lobes. From these values and the morphology of the polarised layer, we infer a super-Alfv\'{e}nic motion of the radio lobes with Mach numbers of $M_A=2.47-3.50$. In addition, we find that the pressure decreases towards regions with higher polarisation degrees. These results favour a magnetic draping scenario instead of a magnetic field compression one.

Since a draped layer suppresses any mixing of the material in the radio lobe with the surrounding environment, but can enhance the mixing and re-acceleration efficiencies inside the lobes, we can explain the flat spectral index. In addition magnetic draping layers are more robust to instabilities at the propagation front.

\section{Outlook}
\label{text_outlook}

Identifying the kind and amount of sources where either magnetic field compression or magnetic draping plays a role has a direct influence on the interpretation of polarisation observations as well as the input for simulations of cosmic-ray propagation and magnetic fields in the intergalactic space. This is even more important for the early Universe. While it is difficult to enrich intergalactic space with regular magnetic fields using starburst type sources, it might be possible with AGN. As we have seen with our results, the morphology of the spectral indices of radio lobes is influenced by the dominant magnetic field morphology and its subsequent processes.

Objects showing such filamentary polarised structures are not necessarily rare, but difficult to detect due to selection effects towards stronger and more compact sources of past radio surveys. Future radio surveys carried out with the new SKA precursor and pathfinder facilities should be able to detect such kinds of diffuse faint sources easily. These instruments not only allow for more detailed studies of sources, like the one we presented in this work, but also observations of sources at higher redshifts. This will allow for the examination of the evolution of magnetic fields over the lifetime of the Universe.

\begin{acknowledgements}
The research leading to these results has received funding from the European Research Council under the European Union's Seventh Framework Programme (FP/2007-2013) / ERC Advanced Grant RADIOLIFE-320745. M. Brienza acknowledges support from INAF under PRIN SKA/CTA 'FORECaST'.  We thank C. Pfrommer for his input to this work. We acknowledge A. Shulevski for providing the 1.4\,GHz radio continuum map and C. Struve for providing the \ion{H}{i} map. We thank R. Beck and D. Worrall for sharing their insight into energy equipartition equations for magnetic fields. The Westerbork Synthesis Radio Telescope is operated by ASTRON (Netherlands Foundation for Research in Astronomy) with support from the Netherlands Foundation for Scientific Research (NWO). This research made use of the Python Kapteyn Package \citepads{KapteynPackage}.
\end{acknowledgements}

\bibliography{/home/bade/Documents/BibTeX/bibtex}{}
\bibliographystyle{aa}

\end{document}